\pdfoutput=1

\documentclass[acmsmall]{acmart}

\AtBeginDocument{%
  \providecommand\BibTeX{{%
    \normalfont B\kern-0.5em{\scshape i\kern-0.25em b}\kern-0.8em\TeX}}}

\setcopyright{acmlicensed}
\acmJournal{PACMHCI}
\acmYear{2021} 
\acmVolume{5} 
\acmNumber{CSCW2}
\acmArticle{395} 
\acmMonth{10} 
\acmPrice{15.00}
\acmDOI{10.1145/3479539}

\usepackage{subfiles}
\usepackage{multirow}
\usepackage{booktabs}
\usepackage{array,ragged2e}

\newcommand{\PreserveBackslash}[1]{\let\temp=\\#1\let\\=\temp}
\newcolumntype{C}[1]{>{\PreserveBackslash\centering}p{#1}}
\newcolumntype{R}[1]{>{\PreserveBackslash\raggedleft}p{#1}}
\newcolumntype{L}[1]{>{\PreserveBackslash\raggedright}p{#1}}

\newcommand{\additionnew}[1]{\textcolor{black}{#1}}
\newcommand{\additionminor}[1]{\textcolor{black}{#1}}

\begin{document}

\title{\additionminor{Designing Transparency Cues in Online News Platforms to Promote Trust: Journalists' \& Consumers' Perspectives}}

\author{Md Momen Bhuiyan}
\affiliation{%
  \institution{Virginia Tech}
  \country{USA}
}
\email{momen@vt.edu}
\author{Hayden Whitley}
\affiliation{%
  \institution{Virginia Tech}
  \country{USA}
}
\email{hwhitley@vt.edu}
\author{Michael Horning}
\affiliation{%
  \institution{Virginia Tech}
  \country{USA}
}
\email{mhorning@vt.edu}
\author{Sang Won Lee}
\affiliation{%
  \institution{Virginia Tech}
  \country{USA}
}
\email{sangwonlee@vt.edu}
\author{Tanushree Mitra}
\authornote{A portion of this work was conducted while the author was at Virginia Tech.}
\affiliation{%
  \institution{University of Washington}
  \country{USA}
}
\email{tmitra@uw.edu}
\renewcommand{\shortauthors}{Md Momen Bhuiyan et. al.}

\begin{abstract}
As news organizations embrace transparency practices on their websites to distinguish themselves from those spreading misinformation, HCI designers have the opportunity to help them effectively utilize the ideals of transparency to build trust. How can we utilize transparency to promote trust in news?
We examine this question through a qualitative lens by interviewing journalists and news consumers---the two stakeholders in a news system.
We designed a scenario to demonstrate transparency features using two fundamental news attributes that convey the trustworthiness of a news article: \emph{source} and \emph{message}. 
In the interviews, our news consumers expressed the idea that news transparency could be best shown by providing indicators of objectivity in two areas (news selection and framing) and by providing indicators of evidence in four areas (presence of source materials, anonymous sourcing, verification, and corrections upon erroneous reporting).
While our journalists agreed with news consumers' suggestions of using evidence indicators, they also suggested additional transparency indicators in areas such as the news reporting process and personal/organizational conflicts of interest.
Prompted by our scenario, participants offered new design considerations for building trustworthy news platforms, such as designing for easy comprehension, presenting appropriate details in news articles (e.g., showing the number and nature of corrections made to an article), and comparing attributes across news organizations to highlight diverging practices.
Comparing the responses from our two stakeholder groups reveals conflicting suggestions with trade-offs between them. 
Our study has implications for HCI designers in building trustworthy news systems.
\end{abstract}



\maketitle

\newcommand{\qt}[2]{{\quote \noindent \small{``\textit{#2}''\hspace{5pt}\textit{--- #1} } \endquote }}

\newcommand{\inlineqt}[2]{\noindent ``\textit{#2}''\textit{--- #1}}

\newcommand{\suggestion}[1]{\hspace{10pt} \textit{\color{blue} #1}}

\newcommand{\centered}[1]{\begin{tabular}{l} #1 \end{tabular}}

\definecolor{OrangeRed}{HTML}{ED3D67}
\definecolor{ForestGreen}{HTML}{000000}

\section{Introduction}

\additionnew{While growth in communication technology has connected news consumers to diverse sources of information, it has also prompted a steady decline in public trust in the mainstream media, as well as the exploitation of public opinion through the spread of ``fake news'' or misinformation~\cite{pew2012further,badawy2018analyzing}.}
\additionnew{
A partial cause of the larger issue of distrust in mainstream media might be attributed to news consumers' inability to distinguish quality journalism from deceptive content; this inability is the focus of this study.
To this end, scholars suggest adopting greater level of transparency within the news-making process~\cite{mcbride2013new,plaisance2009perceptions,seelye2005times}.
}
This shift from a ``trust me'' to a ``show me'' journalism has the potential to promote \additionnew{trust among readers, especially by protecting against erroneous and deceptive sources of reporting~\cite{kovach2014elements,mor2018trust}.}
Online news platforms have the potential to offer novel interface designs to support such an increase in transparency~\cite{coddington2012building,blobaum2014trust}.

\additionnew{Despite calls for transparency in news, 
surprisingly little is known about how to promote transparency to increase trust in news articles, especially from a design perspective.}
In this respect, designers in news organizations need to consider the competing perspectives of stakeholders; for example, journalists may want to protect certain information, whereas news consumers may expect it to be revealed.
\additionnew{Furthermore, journalists and consumers have to consider organizational policy.}
Although some prior scholarly work has explored transparency practices in journalism~\cite{karlsson2010rituals,groenhart2012conceiving,chadha2015newsrooms}, these works discuss them without a particular focus on design and without considering the perspectives of both stakeholder groups.
\additionnew{Some recent work has promoted transparency standards for news websites~\cite{Frontpag7online}, but they focus largely on site-level measures instead of article-level measures.
As discussions to support transparent journalism are just starting to gain traction in HCI communities (see the CHI'19 workshop on the topic~\cite{aitamurto2019hci}),} we have yet to identify effective design principles that can guide the design of these user interfaces.

\additionnew{
This work bridges the gap between communication literature and HCI design in terms of presenting transparency in news articles.
We examine the perspectives of two stakeholders---journalists and news consumers---to inform designers of ways to embed transparency in news articles.
In particular, news organizations both large and small with internal design teams (e.g., The New York Times~\cite{Producta95online}) can benefit from the findings of this work.
Additionally, news aggregators (e.g., Google News) and other third-party news providers (e.g., \texttt{AllSides.com}) may also choose to adopt our design suggestions on how transparency should be practiced in digital news articles.}
Communication scholars have suggested that people take into consideration certain information characteristics when evaluating whether a piece of information is trustworthy. 
People assess the \emph{source} of the message (e.g., the author's expertise) and the \emph{message} itself (e.g., the quality of the message), and they use these insights as meaningful cues to evaluate the overall trustworthiness of a piece of information~\cite{metzger2007making}.
Combining the concept of design cues with transparency, we argue that disclosing certain \emph{source}- and \emph{message}-level transparency cues within news articles can be an important step towards designing trustworthy online news websites~\cite{sundar2008main,Fogg:2003:PTE:765891.765951,karlsson2010rituals}. Overall, we ask:


\vspace{5pt}
\hspace{5pt} RQ. \textit{How can designers \additionnew{utilize transparency cues} to promote trust in digital news articles while considering the perspectives of journalists and news consumers alike?}

\hspace{15pt} RQa. \textit{\additionnew{What aspects of journalistic practice do news consumers and journalists want disclosed within news articles as transparency cues?}}

\hspace{15pt} RQb. \textit{\additionnew{What should designers consider in promoting transparency cues for news consumers and journalists?}}





\vspace{5pt}

\additionnew{To answer these questions, we interviewed journalists and news consumers. Our journalist pool consisted of} 15 journalists with diverse reporting backgrounds who worked in local, national, and international newsrooms. 
For our news consumer interviews, our pool comprised 16 citizens from both local and online communities with novice to savvy news-reading behaviors. 
We complemented our interviews with a scenario-based design approach~\cite{carroll2000making,horning2014scenario}---a user-centered design approach that employs prototype mockups and descriptions to help end-users envision a future system and how they will use it. 
We provided our interview participants with a scenario: a prototype mockup developed with two sets of transparency cues.
These cues disclose underlying journalistic practices pertaining to the \emph{message} being reported and the \emph{source} reporting that message (see figure \ref{fig:newsbeat}). 
Stimulated by our scenario, participants recommended various transparency cues capturing source and message characteristics to address the reasons for underlying distrust, with some additional transparency cues to reveal the process of reporting.
Many of these suggestions concern information which is fundamental to high-quality journalistic news reporting. 

Our analysis revealed that news consumers identified two main areas in which increased transparency could improve trust; namely, the level of objectivity in a report and the quality of the evidence that underlies the claims in a report. 
For objectivity, our analysis showed that designers have the opportunity to incorporate transparency cues to reveal biases in news selection and framing.
For evidence, our news consumers asked for transparency in evidence presentation, anonymous sourcing, fact-checking, and correction.
Meanwhile, our journalists proposed transparency in evidence presentation, the reporting process, and personal/organizational conflicts of interest (e.g., gifts journalists might receive and stakes in other business organizations owners might have). 
Both groups suggested the need for transparency cues to highlight evidence in news reporting, such as by including verification materials. 
The scenario design also prompted participants to point out additional considerations, such as designing for ease of comprehension, reducing bias while presenting transparency indicators, and including appropriate details within the transparency feature (such as displaying both the number and nature of corrections to an article). 
Other suggestions include designing markers for distinguishing reporting quality within news items, and for distinguishing between news and non-news items.
When we compare the responses between our two stakeholder groups, we see conflicting suggestions that would require trade-offs in design. 
For example, trade-offs exist between being fully transparent and preserving autonomy by placing limits on transparency. 
Similar trade-offs exist between opting for simple features to ease comprehension and including sophisticated features to illustrate appropriate nuances. 
We discuss how designers can utilize existing journalistic practices in designing transparency cues catering to our participants' areas of interest, and thus, how they can consider the perspectives of both stakeholders in their designs. 
\additionnew{However, to draw a more complete picture, future work would need to consider news organizations and sources (people) themselves as stakeholders of news. 
Furthermore, while participants responded with consistent themes in the interviews, without properly controlled experiments, our results are not evidence of efficacy of the design ideas. Still,}
our results have implications for design practitioners in highlighting appropriate journalistic practices on news websites through transparency and for \additionnew{design researchers in future investigations of balancing the constraints imposed by stakeholders, considering remaining stakeholders, and conducting controlled experiments to evaluate efficacy.}
\section{Literature Review}

\subsection{Defining Transparency in Journalism}
With rapid improvements in communication technology,
the news-gathering process has shifted from a model of verification (the principal value being getting things right and exercising caution against getting things wrong) to a model of assertion, where the principal value is to get the news out fast, even at the expense of rigorous verification~\cite{karlsson2017not}. 
This shift may result in journalists making more mistakes, thereby directly affecting news consumers' trust in journalism~\cite{arsenault2006conquering}. 
Some scholars argue that transparency (or openness) could be a mechanism for building both trust and accountability~\cite{plaisance2007transparency,allen2008trouble}.
In this work, we define transparency as this notion of openness about \additionnew{the news makers themselves, their} journalistic routines and practices, and their decision-making processes~\cite{tuchman1972objectivity}.
Prior work shows that one of the ways journalists practice transparency is through disclosure of information,
known as \textit{disclosure transparency}~\cite{karlsson2010rituals}.
Disclosure transparency can be realized in different ways, such as how story assignments are decided over meetings, disclosing verification process, linking source material for a report, and acknowledging and correcting errors~\cite{hayes2007shifting,lasica2004transparency}.
When discussing transparency in this work, we utilize this principle of disclosure transparency.


\subsection{Existing Transparency Practices in Journalism}
The earliest examples of transparency practices in journalism date back to the 1930s, when bylines containing the names of reporters responsible for an article were first introduced~\cite{schudson1981discovering}. Practices have evolved from early practices like the use of ombudsman columns (which goes back to the 1960s) to newer ones, such as inviting users to experience the journalistic process firsthand (e.g., attending news meetings)~\cite{craft2009transparency,deuze2003web}.
For a comparison of transparency practices across newsrooms, see~\cite{Discover93online}.
Prior scholarly works looked at various aspects of transparency in online news.
Examining three online newspapers (The New York Times, The Guardian, and Dagens Nyheter), scholars found that each of these organizations implemented different sets of transparency features~\cite{karlsson2010rituals}. 
Revers found wide adoption of transparency in social-media-aided reporting~\cite{revers2014twitterization}.
Offering transparency features across organizations is a comparatively recent phenomenon.
For example, some social media sites have begun offering transparency in news content (e.g., through a corrections and ethics policy)~\cite{HowFaceb12online,Understa80online,Newlabel65online,Designinfeed}.
Despite significant work, research seems to look at transparency in isolation, from either journalists' or news consumers' perspectives alone.
We bridge this gap in existing work by exploring both journalists' and news consumers' views on a problem scenario. 


\subsection{\additionnew{Designing for Trust in News Through Information Disclosure}}\label{sec-rel1}
\additionnew{
HCI research has been considering trust as a driver for design for a considerable amount of time in such areas as e-commerce, remote work, and digital currency~\cite{bossauer2020trust,riegelsberger2003interpersonal,sas2017design,corbett2018going}.
Generally, these works use two concepts of trust: trust being mediated by technology for individual-to-individual relationships and technology being the object of trust~\cite{corritore2003line,gulati2017modelling}.
Existing works support that
openness (or transparency) \additionminor{influences} trust~\cite{knowles2015models}.
In this work, we explore the trust relationship between individuals and information.
In online informational systems, particularly for news, trust is often considered synonymous with credibility~\cite{fogg2003prominence,tseng1999credibility,taneja2019people}, and we adopt this concept in our work.
Communication literature further guides us around the relationship between information disclosure and trust.
}

In communication literature, trust has typically been considered a subset of the larger concept of credibility, where individuals who ``trust'' a given speaker (i.e., the source) or piece of information (the message itself) also deem it more credible~\cite{metzger2007making}. 
Studies have suggested that when people evaluate the trustworthiness of online information, they consider a number of characteristics. These may include evaluations of who conveyed the information (the source), how it was said (the message), and where it was said (medium or channel characteristics)~\cite{kiousis2001public,metzger2007making}. 
However, research also indicates that most people don't spend a tremendous amount of time engaging in close analysis of online content, often relying instead on surface-level features, such as the quality of a website's design or how easy it is to use~\cite{Fogg:2003:PTE:765891.765951}. 
Several theories have expanded this observation and shown that individuals have only a limited number of cognitive resources to evaluate information online; therefore, they often rely on cognitive heuristics to evaluate information~\cite{evans2003two,sundar2008main}. 
By cognitive heuristics, we mean the mental shortcuts individuals use to ignore some information on a site while paying attention to other information when evaluating the trustworthiness of the claims presented. 
On any given website, various design features serve as cues to trigger a heuristic.
For example, 
Facebook uses ``likes'' as a cue to trigger the ``popularity heuristic'' of a given post~\cite{borah2018importance}. 
 
Several researchers have attempted to explain the role that these feature cues play in helping people evaluate information. Fogg's Prominence-Interpretation Theory suggests that people evaluate the trustworthiness of a website based on feature cues that attract users' attention the most~\cite{fogg2003prominence}.
For example, only if a user notices a feature cue (such as an indicator of an author's expertise) would that cue have an impact on the user's overall perception of the source's credibility.
Sundar's MAIN Model also suggests that credibility is conveyed online by four broad website characteristics that serve as their own cues to users for denoting the credibility of information~\cite{sundar2008main}. 

Drawing upon the notion that feature cues can be used as ways to trigger heuristic evaluations of a given piece of content, we explore the design of a series of disclosure transparency cues that aid users in evaluating the trustworthiness of a given piece of online news.
Literature on the subject does not provide any exhaustive list of features for disclosure. 
Therefore, when designing our problem scenario, we selected a few features mentioned in existing literature that could influence users' trust. 
For example, studies have shown that a source's apparent expertise has a significant positive impact on users' trust in it~\cite{meyer2010journalist}.
Similarly, message-level characteristics, such as how crucial pieces of information in an article are emphasized or framed can improve the article's level of transparency and cue readers to evaluate its claims meaningfully. 
In Section~\ref{sec-meth1}, we explain in greater detail how we used these characteristics to develop our scenario.

\subsection{Effect of Transparency on Perception of Trust}
Though scholars suggest a positive association between transparency and perceptions of credibility~\cite{hayes2007shifting,karlsson2011immediacy,kovach2014elements}, when investigated, effects of transparency on news consumers' perceptions of a news article's credibility seem somewhat ambiguous.
While one prior study suggests that a single transparency feature had almost no effect on readers' perception of news credibility~\cite{karlsson2014you}, a recent study supports the idea that multiple transparency indicators can improve an audience's perception of news articles' credibility~\cite{curry2019effects}.
News consumers may also prefer certain features over others (e.g., hyperlinking source material over describing how an article was framed)~\cite{karlsson2018transparency}.
Taken together, there seems to be some effect of transparency on trust; however, proper sets of features must be constructed, and their construction is the focus of this work.

\section{Interviewing News Consumers and Journalists Using a Scenario}
We studied our research question through interviews. For this purpose, we built a scenario to stimulate our participants. Below, we describe the scenario and our recruitment process.

\subsection{Developing a Scenario}\label{sec-meth1}
One of the main challenges in designing transparency attributes in news websites to build trust is obtaining stakeholders' reactions to a system that has not yet been built. In such cases, scholars have suggested opting for an alternate approach, called a scenario-based approach, to facilitate users' reactions~\cite{carroll2000making}. A scenario not only provides concrete examples to stimulate users' responses, but it also provides a holistic view of future possibilities. Furthermore, scenario narratives can be made flexible and adapted to expand a user's imagination. In his seminal work on scenario-based design, Jack Carroll suggests that such an approach motivates a more integrative problem analysis compared to traditional requirement gathering techniques~\cite{rosson2002usability}. We exploited this approach to build an example news article with multiple transparency cues. Our scenario is powered by disclosing two sets of transparency cues: one that captures the characteristics of the \emph{message} being reported, and one that shows the quality and expertise of the \emph{source} reporting that message. 
Source- and message-based cues are fundamental for signaling the underlying credibility of a report~\cite{sundar2008main}, and are thereby key to enhancing users' trust in journalistic practices.  
Figure~\ref{fig:newsbeat} shows the scenario we developed. The left side of this figure contains a news article, and the right side contains the transparency cues. Below, we describe how we adopted and implemented each of the design characteristics shown.

\subsubsection{Source Credibility Cues}
\begin{figure}
    \centering
    \includegraphics[width=0.80\linewidth]{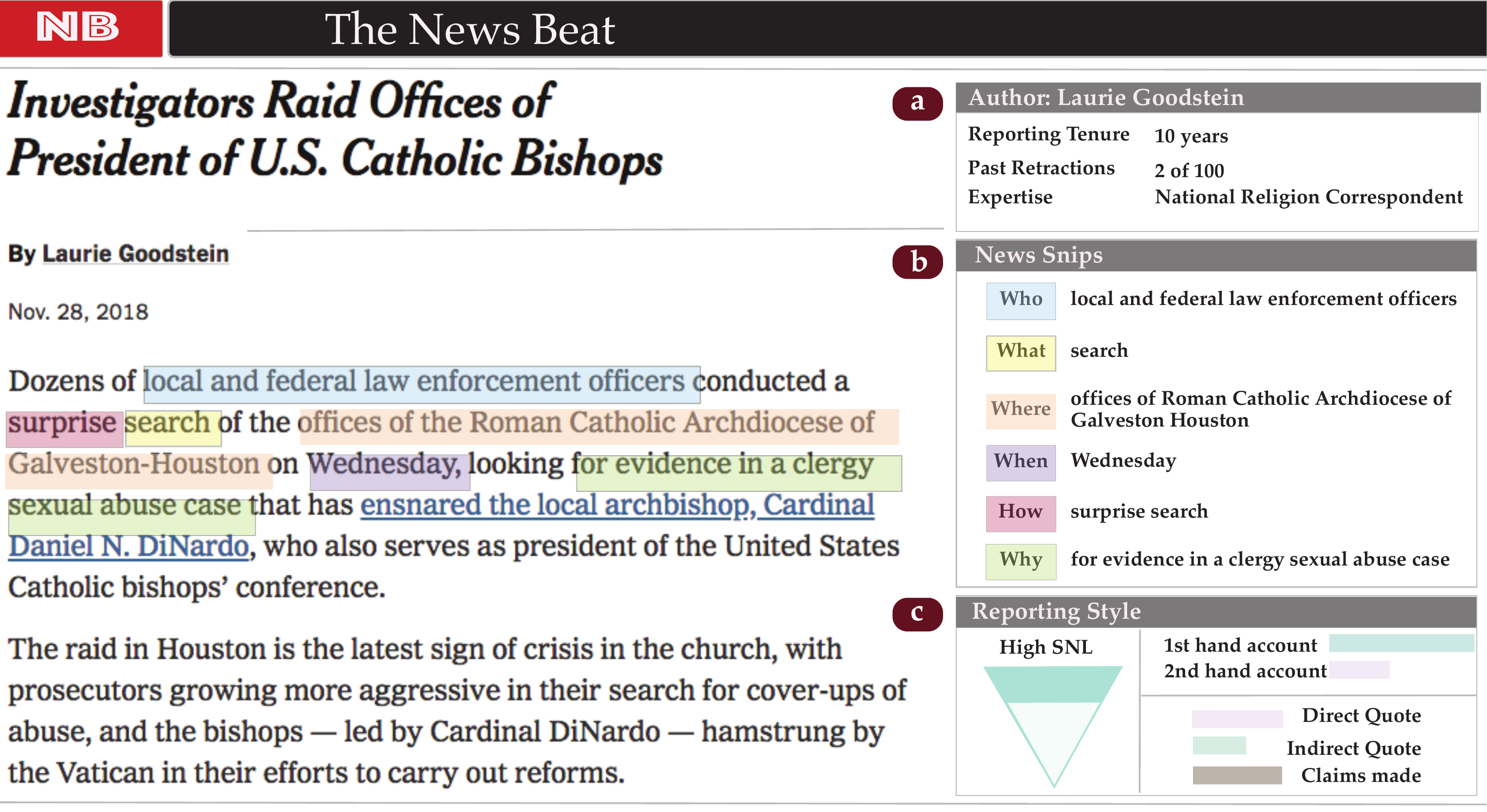}
    \caption{Our scenario with three transparency features. Here, feature ``a'' corresponds to source characteristics conveying the expertise of the author; features ``b'' and ``c'' are message characteristics showing, respectively, crucial details about the event and the reporting style. In feature ``c,'' reporting style includes whether the article is high or low in summary news lead (SNL) or inverted pyramid style reporting, the proportions of first- and secondhand accounts, the proportions of direct and indirect quotes, and the number of claims made.}
    \label{fig:newsbeat}
    \vspace{-8pt}
\end{figure}

Prior literature indicates that source expertise is one of the primary indicators that people use to evaluate the credibility of a news article. In this context, users may consider sources from two perspectives: sources can be taken to be either the organizations that employ journalists or the journalists themselves. In our design, we disclose three indicators of a journalist's expertise, including experience in years, number of retractions, and domain of expertise. Here, we contextualize the number of retractions by presenting it together with the total number of articles written by the journalist. For example, 2/100 indicates that the author had 2 retractions out of the 100 articles she wrote. In Figure~\ref{fig:newsbeat}, we show this information in the upper-right corner as part of feature ``a''. We used this terminology (features ``a,'' ``b,'' and ``c'') during our interviews to simplify references to specific features for our participants.

\subsubsection{Message Credibility Cues}
To give readers cues about the quality of an article, we turn to existing journalistic practices in article writing. We rely on cuing users with the presence of two types of attributes inside a report: attributes identifying the degree of completeness of crucial information in a report (as shown in feature ``b'') and how information is presented within the report in terms of its use of language characteristics that could indicate bias (as shown in feature~``c''). 

\textbf{Cuing the Presence of Crucial Information.} As early as the mid 19th century, traditional journalism has followed the practice of formatting a story by first presenting the crucial information about an event, called main event descriptors~\cite{po2003news}. These main event descriptors answer fundamental questions about an event, and these questions are sometimes referred to as 5W1H: \textit{who, what, where, when, why}, and \emph{how}. To give users cues as to the presence or absence of these descriptors, we highlight each of them in our design with different colors, as shown in Figure~\ref{fig:newsbeat} in the section named ``News Snips'' (or feature ``b''). This feature was inspired partially by Wikipedia's Infoboxes~\cite{HelpInfo85online}, and it isolates the most important information within a news report. These descriptors can also be calculated computationally---multiple prior works have calculated the 5W1H descriptors by applying natural language processing techniques~\cite{wang2008automatic,li2007flexible,hou2015newsminer,wang2012chinese,norambuena2020evaluating}.

\textbf{Cuing Reporting Style.} To give readers cues as to how a report is written, we resort to identifying two aspects of standard journalistic practice in a report: how information is presented and how it is attributed to its source. 
Journalistic practice in structuring news articles varies, though a handful of patterns are commonly used~\cite{harrower2007inside}.
In our design, we focus on ``hard news,'' or news that involves political, economic, and social issues~\cite{limor1997journalism}. This is also a prevalent type of news in the misinformation domain, and these articles tend to follow a particular pattern.
Prior literature suggests that journalists traditionally format hard news by organizing information in a top-down style, starting with the most important information before presenting supporting information in diminishing order of prominence~\cite{po2003news}. This format is also known as the ``inverted pyramid style,'' and it is a common structure in content from news agencies like the Associated Press, whose content is frequently syndicated to many national and international outlets. We examine the degree to which a report follows this standard of writing through a metric called the summary news lead, or SNL~\cite{errico1997evolution}. In Figure~\ref{fig:newsbeat}, the left side of feature ``c'' visualizes this metric as high or low SNL, indicating the degree to which a report follows this style.

Another crucial property of an article is the sources underlying it. Traditionally, journalists divide sources into two categories: firsthand accounts and secondhand accounts. Firsthand accounts are retellings of reporters' direct observations of an event, while secondhand accounts come from authoritative sources with direct knowledge of an event. While both first- and secondhand accounts are useful, firsthand accounts are more accurate~\cite{mencher1997news}. To give users cues as to the nature of sources used in an article, we identify the proportion of claims based on first- and secondhand accounts.

Bias reduction is yet another standard journalism practice. To identify the presence of bias or subjectivity in a report, we examine the framing of quotes in reported secondhand accounts. Reporters can quote secondhand accounts either directly or indirectly~\cite{bergler2006conveying}. They may also introduce their own biases in the process by using framing that alters perspectives~\cite{lin2006side,greene2009more}. For example, a journalist may describe a secondhand account using the word ``claim'' to express doubt in the quoted statement. In prior works, natural language processing has been applied to detect these kinds of cues~\cite{soni2014modeling,sauri2009factbank}. Both observation types and biases are shown to the right of the article as feature ``c'' in Figure~\ref{fig:newsbeat}.

\begin{table}[t]
\centering
{ \sffamily \scriptsize
\begin{tabular}{ccc|ccc|ccc|ccc}
\textbf{Journ.} & \textbf{Gender} & \textbf{Exp. (yrs)} & \textbf{Journ.} & \textbf{Gender} & \textbf{Exp. (yrs)} & \textbf{Journ.} & \textbf{Gender} & \textbf{Exp. (yrs)} & \textbf{Journ.} & \textbf{Gender} & \textbf{Exp. (yrs)} \\ \hline
J1                               & Male                             & 7                                   & J5                               & Female                           & 20                                  & J9                               & Male                             & 5                                   & J13                              & Male                             & 5                                   \\
J2                               & Male                             & 11                                  & J6                               & Male                             & 5                                   & J10                              & Female                           & 9                                   & J14                              & Male                             & 6                                   \\
J3                               & Female                           & 3                                   & J7                               & Female                           & 3                                   & J11                              & Female                           & 7                                   & J15                              & Male                             & 10                                  \\
J4                               & Female                           & 3                                   & J8                               & Male                             & 35                                  & J12                              & Female                           & 3                                   & \multicolumn{1}{l}{}             & \multicolumn{1}{l}{}             & \multicolumn{1}{l}{}\\
\hline
\end{tabular}
}
\caption{Demography of our journalist pool. ``Journ.'' here stands for ``journalist.''}
\label{tab:demo_jour2}
\vspace{-15pt}
\end{table}

\begin{table*}[b]
\centering
\footnotesize
\begin{tabular}{L{0.34\linewidth}L{0.46\linewidth}L{0.13\linewidth}}
 \textbf{Role} & \textbf{Network}  & \textbf{Audience}  \\
\hline
Anchor, editor, reporter (corporate, court, foreign, misinformation, politics), technology research & ABC, AP, Athens Messenger, Buzzfeed, CBC, Daily Reflector, freelance, KBZK, MotherJones, News4SA, Toledo Blade, Roanoke Times, WSJ & Local/metro, national, international \\
\hline
\end{tabular}
\caption{Professional background of our journalist pool. In the network column, ``freelance'' indicates that the journalist is not associated with an organization. \additionnew{Note that we aggregated the roles of journalists and their associations to news networks and audiences to ensure anonymity, as required by the IRB. Knowledge of network affiliation and role would have been enough to reveal the identities of several participants.} }
\label{tab:demo_jour}
\vspace{-15pt}
\end{table*}



\subsection{Recruitment}
Using this design, we interviewed 15 journalists from various news organizations and 16 general news consumers recruited from the local community and online communities on Reddit. Our interviewee pool primarily resided in the United States, with the exception of two international journalists.
We recruited journalists through a combination of personal contacts and social media (Twitter and Facebook).
\additionnew{We recruited a majority of our journalist pool ($n=\additionminor{11}$) from personal contacts.}
\additionnew{Due to the recent focus on misinformation in US politics}, we also \additionnew{searched for journalists covering politics and misinformation on Twitter through keyword searches of} Twitter bios containing a combination of the following terms: ``journalist'' and another word from ``misinformation,'' ``disinformation,'' and ``politics.''
\additionnew{After manually verifying the collected Twitter profiles, we compiled a list and reached out to eight journalists via email.}
\additionminor{Out of the eight, two eventually participated in an interview.}
On Facebook, we joined an invite-only Facebook group of international journalists (IJNet\footnote{https://www.facebook.com/IJNet/}), which comprises a global network of more than 90,000 media professionals.
We posted a recruitment advertisement in this group and were able to recruit 2 international journalists. Through these combined techniques, our final journalist group consisted of 15 journalists, with 2 reporters residing outside the US. 
Our 15 journalists (hereinafter referred as J1--J15) had varying journalism experience, including such roles as anchor, editor, reporter (corporate, court, foreign, misinformation, politics, social media) and technology research.
Their network affiliations included ABC, AP, Athens Messenger, Buzzfeed, CBC, The Daily Reflector, freelance, KBZK, MotherJones, News4SA, Toledo Blade, Roanoke Times, and WSJ.
The distribution of journalists' genders was even. Their experience in the industry ranged from 3 to 35 years, and they served audiences at varying levels, including local/metro ($n=7$), national ($n=6$) and international ($n=2$). Tables~\ref{tab:demo_jour2} and~\ref{tab:demo_jour} show the demography and professional background of our journalist pool.
To preserve journalists' anonymity, we do not report their roles and organizations individually. 


\begin{table*}[t]
\centering
\footnotesize
\begin{tabular}{p{0.12\linewidth}p{0.08\linewidth}p{0.35\linewidth}p{0.16\linewidth} p{0.14\linewidth}}
\textbf{News Cons.} & \textbf{Gender} & \textbf{Profession}           & \textbf{Political Orient.} & \textbf{Leaning} \\
\midrule
U1 & Male & Attorney & Democrat & Conservative\\
U2 & Male & Real estate agent (veteran) & Libertarian & Conservative \\
U3 & Male & Residential service & Republican & Conservative \\
U4 & Male & Educator (retired) & Democrat & Liberal \\
U5 & Female & Director (local theater) & Democrat & Liberal \\
U6 & Male & Construction management (ex-police) & Libertarian & Conservative \\
U7 & Male & Software developer & Independent & Moderate \\
U8 & Female & Graduate student (neuroscience) & Democrat & Moderate \\
U9 & Male & Manager (food service) & Libertarian & Moderate \\
U10 & Male & Undergraduate student (geography) & Independent & Moderate \\
U11 & Male & Undergraduate student (aerospace) & Democrat & Liberal \\
U12 & Female & Financial advisor & -- & Liberal \\
U13 & Female & Management consulting & -- & -- \\
U14 & Female & Special education teacher & --  & -- \\
U15 & Female & Civic organization (ex-nurse) & Independent & Liberal \\
U16* & Female & Civic organization & -- & -- \\
\hline
\end{tabular}
\caption{Demography of our news consumer pool. We asked demographic questions at the beginning of each interview (see Appendix A for the list of questions). Note that some of the participants chose not to specify a political affiliation. * This participant was a former journalist.}
\label{tab:demo_user}
\vspace{-15pt}
\end{table*}
 
For news consumer recruitment, we reached out to the local Chamber of Commerce and several local institutions (e.g., theaters, schools, and civic organizations). 
Additionally, we advertised on Reddit using multiple approaches. We posted on two subreddits dedicated to surveys, {\tt r/favors} and {\tt r/samplesize}. We also sent direct messages to the top 10+ most active members (during the month when we were recruiting) 
in news and politics subreddits, such as {\tt r/politics}, {\tt r/news}, {\tt r/worldnews}, and {\tt r/inthenews}.
We resorted to sending direct messages because the moderators of these subreddits did not allow us to post recruitment advertisements. we reached over 50 users through direct messages sent from a research account. 
We selected the most active members of each community based on the number of comments users in those subreddits posted that month, which we calculated using Bigquery\footnote{\url{https://bigquery.cloud.google.com}}. We also advertised through several mailing lists at our university.
\additionnew{Overall, among our pool of news consumers, we had 2 university students, 2 Reddit users, and 12 participants from local institutions.} 
We offered our news consumers \$15 gift cards for participating in our interviews. 
Table~\ref{tab:demo_user} shows the demography of our pool of news consumers. Considering users' political leanings (the rightmost column), our news consumer pool seemed fairly balanced.
\additionnew{We took political leaning into consideration because it affects users' perceptions of the trustworthiness of various news sources~\cite{USMediaP68online}.}

\subsection{Interview Procedure and Analysis}
\additionnew{Our semi-structured interviews with participants took place mostly over Zoom video call software, though eight interviews with news \additionminor{consumers} took place in person.} A typical interview lasted 40--70 minutes. 
Each interview was recorded and transcribed verbatim by the first author for subsequent analysis.
Using the grounded theory method~\cite{charmaz2006constructing}, the first and second authors open-coded all the interviews and came up with an initial set of themes~\footnote{\additionnew{While some work in CSCW and HCI provides inter-coder reliability for the grounded approach~\cite{mcdonald2019reliability}, it is typically neither appropriate nor required due to a lack of a predefined coding scheme~\cite{burla2008text}}}.
\additionnew{In this phase, these authors read all the transcripts multiple times to determine codes.
Next, these two authors discussed the codes with the other authors to resolve inconsistencies and determine relationships between the codes through axial coding (e.g., combining news selection and framing into objectivity in section~\ref{sec-obj}).
} 
After repeating this process multiple times, we revised the codes into a final set of themes.
We discuss those final themes as they pertain to our research question in the next section.

\additionnew{To summarize, our interviews with the participants (journalists and consumers) revolved around transparency factors related to trust through a series of questions pertaining to reasons for distrust of media (\textit{In your lifetime, do you think the news industry has changed? In what way?}), how emergence of fake news affects news consumption (\textit{As you are probably aware, the subject of fake news is now a popular topic. Has this focus on fake news impacted your news consumption in any way?}), and through a discussion on possible design improvements (after viewing the design scenario, \textit{Would there be any other interactive-type features that you would think could be beneficial?}). See Appendix A for the full list of questions.
For each participant pool, we focused our analysis on two sub-RQs: (RQa) What aspects of journalistic practice do news consumers and journalists want disclosed within news articles as transparency cues? (Sections \ref{sec-con1} and \ref{sec-jour1}) and (RQb) What should designers consider in promoting transparency cues for news consumers and journalists? (Sections \ref{sec-con2} and \ref{sec-jour2}).
However, participants' responses also varied before and after seeing the scenario.
The answers to the sub-RQs are summarized in Table \ref{tab:theme_summary}.
}






\begin{table}[t]
    \centering
    { \sffamily \scriptsize
    \begin{tabular}{L{0.19\textwidth}L{0.10\textwidth}L{0.30\textwidth}L{0.30\textwidth}}
          & \textbf{Pool} & (\textit{Section})  \textbf{Before Showing the Scenario} & (\textit{Section}) \textbf{After Showing the Scenario} \\
         \hline
         \multirow{2}{0.19\textwidth}{RQa. What aspects of journalistic practice do news consumers and journalists want disclosed within news articles as transparency cues?} & News Consumers & \begin{tabular}[t]{p{0.03\textwidth}p{0.25\textwidth}}
             (\ref{sec-obj}) & Transparency in objectivity of news selection and framing 
             \\
             (\ref{sec-evi4}) & Transparency in four aspects of evidence: presentation, sourcing, verification, and correction 
             
         \end{tabular} & \begin{tabular}[t]{p{0.03\textwidth}p{0.25\textwidth}}
            (\ref{sec-auth}) & Transparency in author expertise 
         \end{tabular} \\
         \cline{2-4}
          & Journalists & \begin{tabular}[t]{p{0.03\textwidth}p{0.25\textwidth}}
              (\ref{sec-rep}) & Transparency in  reporting process
         \end{tabular} & \begin{tabular}[t]{p{0.03\textwidth}p{0.25\textwidth}}
              (\ref{sec-evi}) & Transparency in  evidence presentation\\
              (\ref{sec-coi}) & Transparency in conflict of interest 
         \end{tabular} \\
         \hline
         \multirow{2}{0.19\textwidth}{RQb. What should designers consider in promoting transparency cues for news consumers and journalists?} & News Consumers & \begin{tabular}[t]{p{0.03\textwidth}p{0.25\textwidth}}
              (\ref{sec-easy}) & Designing for easy comprehension
         \end{tabular} & \begin{tabular}[t]{p{0.03\textwidth}p{0.25\textwidth}}
              (\ref{sec-hyp}) & Hyperlinking without taking away from the article\\
              (\ref{sec-red}) & Reducing bias considering consumer cynicism 
         \end{tabular} \\
         \cline{2-4}
          & Journalists & \begin{tabular}[t]{p{0.03\textwidth}p{0.25\textwidth}}
              
         \end{tabular} & \begin{tabular}[t]{p{0.03\textwidth}p{0.25\textwidth}}
              (\ref{sec-nua}) & Presenting \additionminor{complicated details} with features \\
              (\ref{sec-cont}) & Contrasting attributes between organizations\\ 
              (\ref{sec-two}) & Two dimensions for distinguishing article quality 
         \end{tabular} \\
         \hline
    \end{tabular}}
    \caption{\additionnew{Theme summary split by when each theme emerged---before or after showing the scenario to the participants. Note that while responses aligned with certain themes emerged both before and after scenario exposure, the themes' positions in this table are dictated by when they emerged most commonly.}}
    \label{tab:theme_summary}
    \vspace{-22pt}
\end{table}


\section{News Consumer Perspective}
Here, we outline results from our news consumers divided into themes that emerged in our analyses. \additionnew{We contextualize them with respect to our two sub-questions regarding attributes to disclose as transparency and design issues to consider.} 
Note that we redacted some entities (e.g., name of a news organization) in our quotes for the sake of anonymity.

\subsection{\additionminor{RQa:} What aspects of journalistic practice do news consumers want disclosed within news articles as transparency cues?}\label{sec-con1}
Analyzing the responses to our interview questions, we found three aspects of reporting regarding source and message to consider for transparency design that may improve trust in news media.
We discuss each aspect below.

\subsubsection{Transparency in objectivity of news selection and framing}\label{sec-obj}
We identified three notions in our interviews that collectively suggest that the lack of trust in the media is partially derived from a perception that the media is losing its ability to be objective in its reporting. Specifically, news consumers highlighted that current practices of heightened polarization, opinion-laden reporting, and sensationalism in headlines have increased mistrust. 
In combination, these themes show a lack of objectivity in two aspects of news reporting: \textit{news selection} and \textit{news framing}.
Here, news selection refers to the process of selecting stories to cover, and news framing refers to how reporters make certain information more salient in an article~\cite{white1950gate,de2005news}.
We present these themes below. 

Many of our news consumers (9/16) emphasized that readers' perceptions of polarization in the media and bias in news stories are two reasons behind distrust of the news. Some of this bias is exhibited in \textbf{news selection}, or what is prioritized in news coverage. News consumers illustrated their view through examples, such as conservatives feeling that left-leaning news organizations prioritize negative coverage of right-leaning politicians and issues, and vice versa.


\qt{U2}{The British tanker gets taken [captured] in the Straits of Hormuz [by Iran]. But that's not the headline. The headline is Trump call Ocasio [a Democratic congresswoman] bad name and told her to go back to wherever she came from.}

\qt{U12}{Like I said earlier, they want to get a certain story. And then they build the facts around what they want, how they want to portray the story, instead of actually just reporting the news.}



 
On top of news selection, news consumers (5/16) often saw news articles loaded with opinion and complained that news organizations \textbf{frame} information based on a particular agenda.
If readers are unable to differentiate between opinion and news, they may run into the pitfall of considering everything opinion and consequently distrusting journalistic news. 

\qt{U5}{And even within the same paper, I think it's easy to lose sight of whether it's an opinion piece or it's an analysis. That's another thing they (news papers) say, ``it's news analysis.'' But the analysis is loaded with opinion.}

\qt{U12}{I'm just giving an example. An earthquake happened. And they show the one building that fell down and say it's destruction, and that's not really what happened.}





Some of our news consumers (4/16) also noted a lack of objectivity in the form of \textbf{unrepresentative headlines}. 
Given the relatively recent influx of alternative news sources, traditional organizations are trying different tactics to seize the attention of consumers. 
Since many people read only headlines, news organizations often attempt to gain consumers' attention by writing headlines that are considered clickbait.

\qt{U7}{Clickbait is a big problem. Even highly reputable sources are resorting to sensationalist headlines at times to grab attention. And obviously, sensationalism has always sold. ``If it bleeds, it leads'' has been a line for decades.}


Taken together, these findings suggest that news consumers are concerned that modern journalism is losing its capacity for objectivity. 
Some see this lack of objectivity in partisan reporting, while others observe it in how reporters choose their words.
Some readers also take issue with headlines that either express a partisan view or engage in some form of trickery to entice users to click and read.
These concerns could be a byproduct of ``agenda setting''\cite{dearing1996agenda}.
Prior research has argued that, as part of ``agenda setting'', news organizations often push a certain narrative by selectively publishing a subset of news stories and focusing on particular pieces of information within a given story~\cite{mccombs1972agenda,entman1993framing,scheufele2007framing}. 
Taking this aspect into consideration, system designers can design cues to distinguish high-quality articles from low-quality articles (discussed in sections~\ref{sec-dis1a} and~\ref{sec-dis1b}).
 
\subsubsection{Transparency in four aspects of evidence: presentation, sourcing, verification, and correction}\label{sec-evi4}
Next to objectivity, our news consumers expressed concerns with how current journalistic practice uses facts and evidence in reporting. Participants focused on four areas relevant to this topic: (1) the presence of evidence details, (2) the use of anonymous sources, (3) decreasing levels of fact-checking, and (4) hard-to-find corrections. Below, we discuss these points regarding evidence.

\textbf{The presence of evidence details} was one important aspect for consumers (4/16) in regard to trusting the legitimacy of news reporting. According to our news consumers, links to source materials and information on all named parties in an article would help users better determine the trustworthiness of an article.

\qt{U10}{Maybe if it has statistics and similar stuff in it... where that information come from? If you know where the source came from, you can kind of test the legitimacy.}

News consumers (5/16) also referred to one particular aspect of sourcing---the rise in \textbf{reports based on anonymous sources}---as a problematic trend in news media and a reason for doubting the trustworthiness of such reports. Some news consumers suggested that much political reporting now relies on anonymous sources, while others pointed out how biases in anonymous sources compel them to doubt the veracity of such reports. 

\qt{U14}{Certainly, anytime anyone quotes an anonymous source, it's a suspect. [Based on] The background of that anonymous source, what biases do they have?}



 
It has long been a common practice in journalism to issue retractions and corrections when journalists have reported something inaccurately. Some news consumers (2/16) expressed concern that modern journalism either seems to \textbf{check facts} less frequently, or that journalists often don't \textbf{correct} themselves when they make mistakes in reporting. These concerns about a lack of visible corrections and retractions could indicate a lack of accountability in reporting.

\qt{U6}{I think the biggest problem with any media or any news now in any format is that when they screw up, they won't own it. They'll dance around and point fingers somewhere else.}

Consequently, news consumers (4/16) mentioned that they were now becoming more reliant on checking facts through other means, such as comparing news stories against related content and using fact-checking services. 

\qt{U8}{I guess, I try to verify things by looking at multiple sources. If a bunch of different sources are all saying the same thing, then that seems more likely to be accurate. And I use websites that are specifically set up for fact checking purposes, like political things.}
 
In this vein, some users (5/16) suggested features to compare journalistic performance. For example, they expressed a need to include information on journalists' past reporting trends, such as hyperlinks to past articles and reviews of the accuracy or political stance present in previous coverage by a given journalist.  

\qt{U9}{Provide a link to maybe a list of their other articles. So, you can see perspectives on them.}





Collectively, our pool of news consumers felt that journalists should present more details on evidence. They expressed concern over the increasing reliance on anonymous source--based reporting and decreasing stress on error correction when journalists make mistakes. They also noted a wider availability of news sources that often complicate ascertaining what is fact. 
These concerns are not unwarranted.
Research shows an increase in anonymous source use in news reports~\cite{purvis2015anonymous,martin2007use} and a significant number (48\%) of factual inaccuracies in US news articles~\cite{porlezza2012news}.
Though there is an increase (twofold, between 1997 and 2007) in the number of corrections over time~\cite{nemeth2009number}, newspapers still fail to correct a large number of mistakes~\cite{maier2007setting,silverman2009regret}.
Transparency regarding each of these aspects can be designed into a news web page (discussed in sections~\ref{sec-dis2a}--\ref{sec-dis2c}). 

\subsubsection{Transparency in author expertise} \label{sec-auth}
After seeing our design, news consumers (5/16) asked for several author-centric details to promote transparency. 
They suggested multiple transparency features, such as outlining the biography of the journalist and including their educational background (\inlineqt{U1}{where they're from, where they went to college}) and other organizations they have worked in (\inlineqt{U7}{... nice to have a way to expand expertise to see all news organization names they worked in.}).
The disclosure of such information could inform consumers about reporters' expertise.
\additionnew{To some news consumers, trusting a journalist is easier than trusting an institution
(\inlineqt{U1}{I don't think you should really trust one news organization. You should look at the reporters that you feel [are trustworthy]}).}
This suggestion is not surprising, given how existing research suggests that expertise is a primary dimension of credibility; background information about a reporter might be a manifestation of such expertise~\cite{metzger2007making}.
In that case, cues that expose the expertise of the organization may also affect news consumers' perceptions of how a story was created, and, in turn, result in greater trust.
Indeed, prior research showed that providing additional information about journalists, such as photos and bios of reporters, fosters greater trust in an organization~\cite{curry2017trust}. 
Our participants provided some nuance to these findings by suggesting specific types of background information, such as journalists' professional and educational background. 
It seems that respondents found most of these suggestions desirable because the information may help individuals determine whether journalists are capable of reporting objectively on a given topic.
For example, consumers might question ``where they're from'' in order to understand how coverage of a story might be affected by the journalist's pre-existing biases.
Considering these suggestions, designers can implement transparency cues that are indicative of reporters' objectivity.
We discuss this later in Section~\ref{sec-dis3}.

\subsection{\additionminor{RQb:} What should designers consider in promoting transparency cues for news consumers?}\label{sec-con2}

\subsubsection{Designing for easy comprehension}\label{sec-easy}
Several news consumers (4/16) preferred information displayed statistically, since it is easy to comprehend. More specifically, they asked for design that presents information in a simplified manner. Oftentimes, they used nontechnical terms to convey this idea.

\qt{U2}{Give me a report, ABCD... like accuracy... like 3/4 key attributes that you're looking for in a reporter [The participant is asking for statistics on a set of easy-to-comprehend evaluation criteria.]}
Having too much information could itself be distracting, as pointed out by some of the news consumers (3/16).
Consequently, this suggestion might be coming from a tendency to reduce cognitive load.
Taking these problems into consideration, designers could implement transparency features as simple markers and give users the ability to switch them on and off.
 
\subsubsection{Hyperlinking without taking away from the report}\label{sec-hyp}
Though several news consumers (5/16) suggested hyperlinking as a way to provide more information, some (2/16) insisted that hyperlinks take users away from the primary news and might thus distract readers.
\qt{U8}{If there is even one more click to get to that information [more transparency detail]... I feel like losing tons of people. Because we're all just so lazy or busy.}

This suggestion to reduce distractions might be another reference to reducing cognitive load in getting information. To address this problem, web pages could provide previews of each hyperlink whenever a user hovers over them. For example, Wikipedia has already implemented such a preview feature to provide context without the cost of context switching~\cite{Howwedes11online}. 

\subsubsection{Reducing bias considering consumer cynicism}\label{sec-red}
Throughout the interviews, our news consumers (7/16) showed cynicism towards news media in general.
Referring to biases in news reporting, news consumers expressed a reluctance to trust any information.
Consequently, some of them (3/16) suggested that transparency features should come from a third-party reviewer.

\qt{U12}{I still watch the news. I want to be informed. But I don't take everything for face value.}
\qt{U7}{Potentially not even in their own words, in a third party words... like an outside reviewer that judges their way of what they've written previously and puts them on a political scale.}

This suggestion for outside oversight in transparency indicators implies that news consumers suspect news organizations of making improper claims of transparency.
If readers cannot trust the transparency indicators, they will have no effect.
To design against cynicism, designers may apply de-biasing techniques to modify either the environment or the decision-maker~\cite{nagtegaal2020designing}; in our case, these are the design and the news consumer, respectively.
As participant U7 mentioned, setting up a (bipartisan) third-party authority might be useful as a modification to the environment. 
Alternatively, designers can pursue nudging techniques, such as considering the opposite (e.g., asking news consumers why a statistic is inappropriate)~\cite{lord1984considering,adame2016training}.
 
\section{Journalist Perspective}
\additionnew{In this section, we outline the themes from journalists as we did with those from news consumers with respect to our sub--research questions about attributes to disclose for transparency and design issues to consider.}

\subsection{\additionminor{RQa:} What aspects of journalistic practice do journalists want disclosed within news articles as transparency cues?}\label{sec-jour1}
Our journalists suggested transparency in two specific areas: characteristics of news organizations and the reporting process. Overall, they discussed three features for transparency: (i) providing evidence to support an article, (ii) emphasizing the process of reporting, and (iii) noting the reporter or organization's conflicts of interest.
Below, we elaborate on their definition of trust and the areas where they suggested increased transparency.
 
\subsubsection{Transparency in evidence presentation}\label{sec-evi}
A majority of our journalists (10/15) brought up evidence as one key area for improving trust in news. Some of them (5/15) suggested that publishing unedited evidence material alongside a report can help news consumers to fact-check the report. We also found that news consumers preferred such transparency around evidence.


\qt{J13}{I would say... give out documents for open access. I am for opening that up so that people can check it up... some leverage for the people to do fact checking on the journalist's work.}


Providing source material can empower news consumers to verify information themselves, thus streamlining their capability to fact-check while holding journalists more accountable.
For example, transparency cues that highlight a position (e.g., a page number or time code) within a source (e.g., a document, audio recording, or video) can help news consumers identify discrepancies between the source material and an article based on it.
Studies have found that the use of hyperlinks for evidence on news sites can increase perceptions of news credibility, drive users to seek out more information, and drive longer engagement~\cite{weber2012newspapers}.
We further discuss how designers can utilize this suggestion in section~\ref{sec-dis2a}.
 
\subsubsection{Transparency in the reporting process}\label{sec-rep}
Our scenario suggestions also prompted several journalists (4/15) to think about other ways newsrooms could show news audiences their reporting processes, such as by providing behind-the-scenes details. Such contextualization could include how journalists went about making decisions while writing a story, thus revealing the underlying journalistic process of news reporting.

\qt{J9}{I think as soon as the meeting ends, I, as a reporter, can go to Facebook or social media, and say, this is what I'm doing today, this is what I'm doing right away at 9 o'clock in the morning, this is why I do this. what do y'all think about it?}

Journalists' suggestions to show the processes that produce their reports seem to be an attempt to show professional practices in their newsrooms compared to others, and to allow consumers to compare the quality of the journalistic standards of various organizations. 
Some research suggests that explaining elements of the reporting process, such as verification procedures, increases news consumers' trust in a given article~\cite{Research95online}.
Research also points out that compared to author-related attributes, such as a journalist's biography, showing evidence or a behind-the-scenes verification process more significantly enhances users' trust~\cite{curry2017trust}.
We discuss what transparency cues designers can utilize regarding behind-the-scenes verification in section~\ref{sec-dis4}. 

\subsubsection{Transparency in conflicts of interest}\label{sec-coi}
Apart from the features related to author expertise and corrections provided in our scenario, several journalists (4/15) suggested conflicts of interest as a consideration for transparency. They suggested that both journalists and organizations can be more transparent in presenting their conflicts of interest, such as relationships with the people that journalists are reporting on (\inlineqt{J12}{... reporting mechanism for meetings when you go to different parties and lobbying events.}) or the financial background of a journalistic organization (\inlineqt{J2}{The key would be if it is a ... for-profit/non-profit organization. If for-profit, is there an easy way to to show profitability? funding-wise where it is coming from.}). Such details may contextualize the perspective in a report and help consumers identify any potential bias in coverage. We further discuss how designers can follow existing practices to show transparency on conflicts of interest in section~\ref{sec-dis5}.

\subsection{\additionminor{RQb:} What should designers consider in promoting transparency cues for journalists?}\label{sec-jour2}
\subsubsection{Presenting \additionminor{complicated details} with features}\label{sec-nua}
After reviewing our scenario with several transparency features (e.g., years in reporting, retractions), a majority of the journalists (8/15) raised concerns about missing complicated details in our design.
For example, our journalists suggested that reporting the number of years a reporter has worked in journalism does not necessarily reflect whether the reporter has done high-quality work throughout that time. Conversely, readers could use this metric to discount high-quality coverage from less experienced journalists:

\qt{J1}{It's like... first time I am getting in the news business... writing a story... reporting my first story... and somebody immediately rejects it because they see that the person only had one year of professional [journalistic] experience.}

Several journalists (4/15) similarly suggested improving the retraction metric by including additional disclosures as to the nature of retractions and corrections, ranging from ``small mistakes,'' like minor technicalities or spelling errors, to ``severe mistakes,'' like factual errors.

\qt{J14}{Those two things (corrections) could have been the address of a building for example, which is meaningless. Or they could have been like just fundamental facts about a story. Like she said, x happened and it didn't happen.}

These suggestions imply that designers may inadvertently create a false sense of accuracy if they do not lay out these additional nuances. To address this issue, designers need to handle these details appropriately, with statistical evidence.

\subsubsection{Contrasting attributes between organizations}\label{sec-cont}
While discussing our feature set in the scenario, several journalists (4/16) suggested that some transparency features (e.g., corrections) regarding journalistic practices might not be comparable between organizations.
Though designers can promote transparency to compare practices within and between organizations, our journalists were particularly concerned about inter-organization comparisons.
They reasoned that organizations often vary in their practices and standards, making it harder to conceptualize a fair comparison of such practices.
For example, referring to the number of retractions in the scenario, some journalists mentioned diverging practices across organizations: Some organizations might issue correction in an article for minor errors, while others may not.
Consequently, such comparisons could lead readers to draw mistaken conclusions.

\qt{J10}{It depends on the situation. [For minor framing issues,] some news organization might change that wording but not issue a correction. Others will issue a correction and change the wording. Others won't do anything at all.}

Due to lack of standards in practices across organizations, designers may seek to create their own standards for presenting this contrast. For instance, they could set criteria for retractions and penalize institutions that do not follow them. 
If designers apply such methods to standardize these practices, news consumers should be informed about the standards.

\subsubsection{Two dimensions for distinguishing report quality}\label{sec-two}
From their domain knowledge, our journalists (4/15) showed interest in differentiating the quality of a journalistic report along two dimensions: (i) quality within news items and (ii) quality between news and non-news items.
With respect to differentiating quality within a report, our journalists suggested markers to signal original and derivative work (\inlineqt{J2}{...that's good because it will show if it's an original piece of work, as opposed to a derivative.}) and markers to denote whether the story is a breaking news report (\inlineqt{J4}{So a pop-up-like disclaimer that this is what we're doing right now, we are trying to collect the information.}, speaking about a piece of breaking news).
For differentiating along the news/non-news dimension, journalists suggested markers that indicate news versus opinion, news versus satire (\inlineqt{J12}{...maybe something that scrapes the website, looks at the `about me,' looks for satire and flags that.}), and credible news versus misinformation (\inlineqt{J2}{has it been, like, highly distributed and shared across channels that are sort of questionable?}).
Designers need to make news consumers aware of the relationship between each attribute and the corresponding dimension.
Considering news consumers' desire for simplicity, designers may seek to prioritize one dimension over others.
For example, designers might consider distinguishing news versus opinion, as research indicates that the majority of US adults are poor at differentiating between the two~\cite{CanAmeri14online}.

\section{Discussion}\label{sec-conjour}
Our news consumers and journalists suggested several aspects of news coverage where transparency cues can be helpful, as well as a set of design issues to consider.
\additionnew{Below, we first compare the two groups (section~\ref{sec-comp}), followed by discussions of how designers can leverage aspects of transparency in existing organizational practices (section~\ref{sec-util}) and of design issues (section~\ref{sec-issues}).}

\subsection{Comparing the Perspectives of News Consumers and Journalists}\label{sec-comp}



\subsubsection{Agreement on presenting evidence and differences in reporting aspects for transparency}
News consumers and journalists alike agreed that presenting evidence details can effectively improve transparency in news reporting. 
In terms of disclosing the reporting context for transparency, our analysis shows that news consumers focus on a given report with only passing interest in the reporter, while journalists' suggestions encompassed four main areas: the report, the reporter, the reporting organization, and the reporting process. 
While our news consumers were often interested in indicators of bias in a story, our journalists showed openness to sharing information, such as revealing conflicts of interest and reporting procedures to address concerns around bias.
 
\subsubsection{Conflict between simplicity and nuance in design}
Recall that several news consumers asked for designs that simplify information. Contrary to that, our journalists' objections to some of our design features also point out concerns that simple metrics may be insufficient without additional context.
From a technological standpoint, our journalist pool seemed more savvy and provided technical details (e.g., ``heatmap'' and ``pop-up'') when discussing feature designs.
Comparatively, some of our news consumers offered suggestions in more general terms (e.g., ``accuracy'').
Given this tension between the two stakeholder groups, designers would have to be conscious of trade-offs between simplicity and nuance in the design of these features. For example, they may provide specific, nuanced information in some cases (e.g., severity of corrections) while excluding other details.
When such a level of nuance is required, designers can engage the audience by adding summary markers as a means of digging deeper into more general statistics.
 

\subsubsection{Conflict between transparency and autonomy}\label{sec-conf-ta}
Several journalists (5/15) felt that some of the transparency features were impractical due to ethical, professional, or corporate boundaries. Some mentioned their own stances on protecting the confidentiality of sources, especially anonymous sources like whistleblowers, fearing backlash. The pervasiveness of anonymous sources in political reporting may reinforce this stance. Others mentioned that news organizations were unlikely to support some of the suggestions, such as televising meetings, from fear of exposing trade secrets. 
\additionnew{However, there have also been instances in the past where an organization, such as the NYT, televised its editorial meetings~\cite{TimesCas56online}. These instances suggest that organizations could be open to these kinds of transparency practices.}

\qt{J9}{Will that [putting a live camera in a morning meeting] ever happen? I don't think so. Because there's just a lot of stuff that we talked about in that meeting that's sort of, like, behind-the-curtain stuff. And we talk about like, should we do the story about the school? Yeah! Because our demo[graphic] is 24- to 44-year-old moms who care about the school system.}

Prior research suggests that transparency, which creates a limited form of accountability~\cite{fox2007uncertain}, can facilitate severe scrutiny and restrict journalistic autonomy~\cite{allen2008trouble}. 
As news consumers seek greater disclosure, organizations might impose constraints on full transparency due to concerns regarding secrecy and maintaining an autonomous corporate image.
Designers will have to address this tension between stakeholders that arises from these competing values. 
\additionnew{
To address this issue, designers could ask at least three questions when implementing a transparency feature in a news article: \textit{Does this feature violate any policy of the organization? Does this feature violate the privacy expectations of a source referenced in the article? Do news consumers desire this transparency feature?} If the answer to either of the first two questions are yes, designers would have to revise the feature.
To this end, understanding the policy norms of news organizations, as well as the privacy expectations of sources referenced in a news report, might be an important consideration for research.}

\subsection{Design Suggestions Based on Existing Journalistic Practices}\label{sec-util}
Both of our participant groups suggested several aspects of transparency for reporting. What design cues can HCI designers build around these aspects? We propose that these cues be developed on the basis of existing practices in news organizations. 
\additionnew{These design suggestions are summarized in Table~\ref{tab:design_suggestions}.}

\subsubsection{Newsworthiness cues for news selection bias}\label{sec-dis1a}
Considering objectivity, our news consumers referred to two areas where bias is injected in the production of news: news selection and framing.
Journalists select stories to cover using criteria for \textit{newsworthiness}, also known as \textit{news values}, based on their desire to appeal to public interests~\cite{burns2018understanding}.
Though early research proposed such news values as timeliness (how recently the event occurred), proximity (how close to the audience the event took place), conflict, and sensationalism~\cite{galtung1965structure,shoemaker1987deviance}, journalism has evolved to consider additional news values, such as eliteness (presence of an entity with great societal power), exclusivity, and entertainment~\cite{harcup2017news}.
Empirically, scholars have found that conflict and eliteness are the strongest predictors of newsworthiness~\cite{boukes2020newsworthiness}.
Existing transparency practices in the media often broadly specify news selection on a site-level basis. For example, ProPublica offers a mission statement that says ``to expose abuses of power...''~\cite{ThisIsWh21online}.
Compared to such site-level practices, designers could offer transparency cues explaining the newsworthiness of each article.
For example, they can identify the criteria of newsworthiness that a particular story meets.
 
\subsubsection{Fairness cues for framing bias}\label{sec-dis1b}
The fairness concern is studied in communication literature as \textit{framing bias} (i. e., levels of opinion within a given report).
Scholars propose that journalists raise the salience of specific information to prime the target audience into thinking in a particular way~\cite{entman1993framing,nisbet2002knowledge}. 
Such framing is often found in coverage of congressional candidates~\cite{druckman2005impact,Sides_2016}, tax policies~\cite{entman2005constraining} and racial issues~\cite{gandy1997race}.
As one news consumer suggested, left-leaning organizations often cover right-leaning politicians by framing their point of view negatively.
Research also suggests that journalists can be unaware of their own biases in how they select certain words, and in how they omit or decide to include certain details~\cite{Kuypers_2002,Kuypers_2013}.
Overall, prior research indicates that a skew in fairness exists in the coverage of certain news areas.
To differentiate balanced news stories from ones that are skewed, designers can promote design cues that indicate the viewpoints or sources covered in a story and how such viewpoints are presented. 
For example, designers can use computational tools to identify all named parties (both persons and organizations) in a report and detect the author's or organization's stance towards each of them~\cite{augenstein2016stance,lample2016neural}.
Transparency cues could also make consumers aware of the efforts of journalists to get multiple sides of a given story.
Journalists might consider disclosing which viewpoint(s) they were unable to cover or did not receive any comment on, as well as the extent of their efforts to obtain such information. 



\subsubsection{Presence of Evidence Cues}\label{sec-dis2a}
Our news consumers suggested opportunities for transparency cues that highlight good journalistic practices relating to source materials.
Designers can construct transparency cues to indicate how sources are presented in a report. 
For example, design cues can differentiate the existence of source materials, show the timeline of the collection of such materials, and indicate how information from the source materials is presented in a given report.
In this respect, designers can utilize computational tools to highlight evidence details.
For instance, they might indicate whether all named parties in a report have hyperlinks, whether sources are referenced ambiguously (e.g., ``sources said'' vs ``<name>, a spokesperson for <company>, said''), and contextualize quotes (e.g., when users hover over a quote, a tooltip can show the full paragraph containing the quote).

\subsubsection{Anonymous Source Cues}\label{sec-dis2b}
A recent survey suggests that although news consumers realize the numerous reasons for using anonymous sources, they are concerned that news organizations unnecessarily omit justification for this practice~\cite{WhatAmer37online}.
To justify their use according to the Associated Press's 2014 guidelines~\cite{associated2014ap}, designers may show transparency cues regarding why the requested information is not available on the record, how reliable the source is, and what resources (e.g., public documents, on-the-record sources, and reactions from those affected by a story) have been used to corroborate information from anonymous sources. 
In practice, some organizations show their verification procedures for anonymous sources at the site level, instead of at the article level.
For example, ProPublica shares how they generally verify anonymous sources without going into specifics for each case~\cite{HowDoWeV62online}.
Taking a step further, designers can establish verification standards for anonymous reporting and provide design cues outlining them on news websites.
For example, designers can show the degree of acceptability of various verification materials (e.g., public documents compared to reactions from those affected by a story).

\begin{table}[t]
    \centering
    { \sffamily \scriptsize
    \begin{tabular}{R{.12\textwidth}L{.41\textwidth}L{.2\textwidth}L{.08\textwidth} L{.08\textwidth}}
         \textbf{Transparency Cues (Section)} & \textbf{Example Questions} & \textbf{Implementation Requirements}  & \textbf{Suggested By} & \textbf{Disagreed By} \\
         \toprule
         Newsworthiness Cues (6.2.1) & 
              Which news values does this report represent (e.g., conflict, sensationalism, eliteness and entertainment)? To what degree?
          &  \additionminor{Requires access to organizations' news values} & News consumers & - \\
          \hline
         Fairness Cues (6.2.2) & 
              Who are the named parties in this article?
              To what degree does this report represent each political affiliation (left/center/right)?
              Did the reporter receive comments from all contacted parties?
          & \additionminor{Requires knowledge of organizations' procedures for getting comments} & News consumers & - \\ \hline
         Presence of Evidence Cue (6.2.3) & 
               Does the report cite an authoritative source of evidence?
               Is there ambiguity in how sources are represented?
          & \additionminor{Requires access to official source materials (might be openly available)} & Both groups & - \\ \hline
         Anonymous Source Cue (6.2.4) & 
               Does this report contain anonymously sourced information? 
               Why was the information not available without anonymity?
               How did the reporter verify the information?
               How acceptable is the verification material?
          & \additionminor{Requires knowledge of organizations' procedures for anonymous sourcing} & News consumers & Some journalists\\ \hline
         Fact-check Cue (6.2.5) & 
               Has any internal/external entity fact-checked this information?
               Who fact-checked it (with links)?
          & \additionminor{Requires access to internal/external fact checkers and their procedures} & Both groups & - \\ \hline
         Correction Cue (6.2.5) & 
               Have there been any corrections to this report? Why?
               How were the corrections framed? 
          & \additionminor{Requires knowledge of organizations' correction protocols} & Both groups & Some journalists \\ \hline
         Author Expertise Cue (6.2.6) & 
               What skills does the reporter have?
               What is the reporter's educational background?
               How objective has the reporter been in past reports?
          & \additionminor{Requires comprehensive knowledge of journalists' reporting history} & Both groups & Some journalists \\ \hline
         Behind-the-scenes Cue (6.2.7) & 
              Does this report contain any behind-the-scenes details?
              How did the reporting process occur over time?
          & \additionminor{Requires access to behind-the-scenes materials for a report} & Journalists & Some journalists \\ \hline
         Conflict of Interest Cue (6.2.8) & 
               Does this report cover any entity with which the reporter/organization has a conflict of interest?
               How does the reporter/organization deal with such conflicts?
         & \additionminor{Requires access to news organizations'/journalists' financial information} & Journalists & - \\ 
         \bottomrule
    \end{tabular}}
    \caption{Design suggestions summarized according to two criteria: implementation requirements and consensus or disagreement among participants. Here, the requirements of access to (and knowledge of) an organization's resources (and protocols), such as internal/external databases of prior corrections, conflicts of interest, and behind-the-scenes materials could make it difficult for a third party to implement the design cues. The two right-most column suggest both within-group and between-group disagreement among our participants. 
    As an example, the Behind-the-scenes Cue, discussed in section~\ref{sec-dis4}, could be hard to implement without access to organizations' materials. Some of the journalists disagreed as to the feasibility of disclosing portions of this information (e.g., televising meetings).
    \additionminor{Another example is the Author Expertise Cue (section~\ref{sec-dis3}) discussed by both groups with some disagreement from journalists due to its (e.g., years in reporting) negative impact on new journalists, which could be hard to implement due to the required access to organizations' protocols.} Additionally, we provided sample questions designers could use to build transparency cues.}
    \label{tab:design_suggestions}
    \vspace{-22pt}
\end{table}

\subsubsection{Fact-checking and Correction Cues}\label{sec-dis2c}
Considering the concerns pertaining to fact-checking and corrections, designers could offer transparency cues that detail who fact-checked the information (e.g., reporter / copy editor / editor / third-party fact-checker), quick links to ask ombudsmen to verify the information, a timeline of changes to the article, and statistics pertaining to corrections made by the organization and the author.
Computationally, designers can also use third-party trackers (e.g., \url{www.newsdiffs.org}) that analyze changes in news reports published by an organization and show what changes were made.
Designers could propose cues that highlight how corrections are framed in a report, signifying the degree to which organizations take responsibility for errors in their coverage.
For example, when news media organizations include corrections, they often use a range of framing devices, such as ``clarification'' (evades responsibility) and ``apology'' (assumes responsibility)~\cite{gilboa2019errors}.
 
\subsubsection{Author Expertise Cues}\label{sec-dis3}
In addition to article-related aspects, our news consumers suggested several features related to author expertise.
In practice, some organizations display biographies of their journalists along with their professional histories (e.g., reporting areas) and personal information (e.g., education) on their websites.
Designers could additionally indicate differences, such as desired values in journalism.
Existing surveys list several characteristics (e.g., skills, knowledge, and work values) and their importance in the journalism profession~\cite{2730220041online}.
Using this information, designers can prioritize showing certain information (e.g., active listening skills, which are more important than time management skills).
However, as suggested in the interviews, designers would need to consider nonstandard practices across organizations before showing these contrasts.
Comparisons within an organization might also not be meaningful in some cases, due to skewed distributions (e.g., the level of education for New York Times reporters~\cite{wai2018expertise}).
Moreover, comparisons between generations can be tricky due to changes in required skills (e.g., the addition of technological skills in current job postings~\cite{2730220041online}).
Apart from professional skills, HCI designers can also focus on journalists' expertise in being objective, since some news consumers positioned author expertise in relation to objectivity in a report.
Transparency in journalists' and their editors' writing skills might be useful in this regard.
Furthermore, designers can construct new indicators, such as summaries of fairness in journalists' past reports, signaling their historical patterns of objectivity.

\subsubsection{Behind-the-scenes Cues for the Reporting Process}\label{sec-dis4}
Apart from presenting ample evidence in the report, our journalists proposed greater disclosure of reporting processes or behind-the-scenes details as another avenue of transparency.
In the past, some organizations have used such practices as inviting community members to meetings to disclose these details.
Some recent examples in which behind-the-scenes details are disclosed include podcasts on how investigative journalists track their sources, and tweets by journalists containing their daily notebooks~\cite{SeasonOn95online,HowoneWa71online}.
Designers can adopt these details when creating news websites.
For example, by leveraging diverse sources of information (e.g., social media posts and organizations' internal trackers), designers can present details that correspond to a given report in a timeline.
Furthermore, HCI designers might also consider building systems to support journalists in sharing their work in progress and collaborating with the community~\cite{25WaysCo91online}.

\subsubsection{Conflict of Interest Cues}\label{sec-dis5}
Transparently disclosing an organization's conflicts of interest may be useful in providing news consumers with insight into the financial holdings of a company, which might impact its reporting.
For example, many consumers aren't aware that ABC is owned by The Walt Disney Company, while CBS is owned by ViacomCBS, a subsidiary of National Amusements. McChesney has extensively documented the conflicts of interest that arise with news organizations when they are forced to cover issues that impact their parent companies~\cite{mcchesney2004problem}. 
While currently not prevalent in journalism, 
some organizations engage in practices detailing funding sources at the site level.
For example, several organizations, including The Economist~\cite{AboutusT46online}, NJ Spotlight~\cite{FundersN0online}, and ProPublica~\cite{Supporte60online}) list all of their donors.
Some others, such as The Wisconsin Center for Investigative journalism~\cite{FundingW20online}, provide extensive details that include donation amounts. 
However, a handful go so far as to highlight when a donor is mentioned in a story (e.g., Texas Tribune). 
Informed by existing practices, designers could promote design cues at the article level whenever donors are mentioned.

\subsection{Considering Design Issues}\label{sec-issues}

\subsubsection{Conflicting priorities: value-sensitive design}
Throughout the study, we noticed several conflicts between our two stakeholder groups.
For example, news consumers prefer designs that are quick and easy to comprehend, while journalists prefer designs that provide sufficient nuance in transparency attributes. Similarly, journalists would like to retain their autonomy, while news consumers want more accountability on the organizational level.
To address these conflicts, HCI designers can utilize an existing design approach known as \textit{value-sensitive design}~\cite{friedman1996value}.
This approach suggests that designers should not consider conflicts as ``either/or'' situations, but rather as constraints on the design space~\cite{friedman2008value}.
Depending on the nature of the conflict, designers might be able to balance some tension using existing HCI design principles.
For example, \textit{social translucence} is a design principle that addresses the accountability-autonomy conflict~\cite{erickson2000social}.
To illustrate, in the case of putting live cameras in news meetings, designers could incorporate \textit{social translucence} by showing the video while redacting sensitive content through appropriate methods, such as blurring and muting.
Such details could lead to increased visibility of journalists' activity while imposing limited accountability on the journalists' work, without sacrificing privacy.
When balance between conflicts is not feasible, designers could discuss a workable design space with the stakeholders.
For example, news consumers may spare transparency in some aspects (e.g., the identity of an anonymous source) while requiring it in others (e.g., how many times a particular anonymous source was used).

\subsubsection{Organizations' openness to transparency}
\additionnew{Some of our participants were wary of whether organizations would be open to implementing site-level features (e.g., \inlineqt{U8}{I don't foresee it being adopted broadly by news organizations voluntarily. Because I think the commercial ones, especially, would see it as just a cost with no benefit.}).
Prior scholarly work has revealed organizations' unwillingness to increase transparency regarding certain aspects, including their methods and motives~\cite{chadha2015newsrooms,kovach2014elements}.
However, there has also been a general trend of news organizations adopting greater transparency.
In this respect, some news organizations have been opting for an in-house, ad hoc framework of transparency.
For example, Axios created a ``bill of rights'' for news production that prioritizes transparency surrounding editorial ethics policies for sources and article corrections~\cite{AxiosEdi7online}.
Meanwhile, a large number of national and international organizations ($n>200$) are adopting third-party frameworks on transparency (e.g., Trust Project)~\cite{Frontpag7online}.
In its 37 transparency indicators, the Trust Project largely focuses on site-level features, with only a handful of article-level transparency indicators, such as disclosing corrections and distinguishing between news and other kinds of material (viz., opinion, satire, advertising)~\cite{Collabor30online}. 
Comparatively, our work illuminates ways in which listed organizations can transform site-level features into more specific, article-level features.
For example, as mentioned earlier, instead of providing a comprehensive list of donors, designers can offer transparency whenever a donor is mentioned in a report.
Article-level transparency may especially help with the concern that news consumers may not notice transparency features~\cite{Maybegre31online}.
In general, this work offers guidance for priorities, as well as design issues to consider for transparency.
}

\subsubsection{Standardizing organizational practices for contrast}
Our discussion with journalists emphasizes the need for a set of transparency metrics that can be compared across news organizations.
A precursor for such an implementation is a standardization of practices across organizations. 
If consensus develops around a standard that is common to a group of stakeholders, it may guide organizations to follow that standard when reporting news.
Considering recent efforts in standardizing transparency practices (e.g., the Trust Project), creating credibility standards (e.g., W3C Credible Web Community Group~\cite{Credible12online}), and creating transparency markup standards (\url{schema.org}\cite{Markupfo36online}), standards in organizational practice assist can these efforts.
Notably, past platforms like news aggregators had to work with external websites and tools (e.g., Politifact and the Share the Facts widget~\cite{Sharethe17online}) to identify fact checks.
Designers can address this problem by transmitting transparency details as metadata and making it easier for site designers across websites to show transparency without added hassle.
In addition, standardization of organizational practices could create further barriers to fringe sources' constant production of false stories. 
For instance, if providing source material in a report is mandatory per transparency standards, it may place a burden on the creators and maintainers of alternative and fake news websites, therefore slowing their pace of publication.

\subsubsection{Feasibility of transparency cues in practice}
\additionnew{In addition to the complexity arising from the lack of standards in practices among organizations, designers would also have to comply with the priorities of the organizations for which they work.
As we mentioned before, apart from dedicated design teams in news organizations (e.g., the design team at the NYT~\cite{Producta95online}), other third-party news entities could also adopt our results
by providing ``transparency as a service'' through either a browser extension or a dedicated website.
For example, \texttt{Allsides.com} boasts of showing news from the left-center-right perspective~\cite{AllSides1online}.
Here, \texttt{Allsides.com} may have a priority in being transparent about biases in left-center-right news sources.
For such third-party entities, designers may have to consider their access to organizational resources in implementing certain features (e.g., prior reporting history).
To this end, we summarized the implementation feasibility for our design suggestions based on access to an organization's resources in Table~\ref{tab:design_suggestions}.
Here, behind-the-scenes details are only available directly from organizations, while cues for presence of evidence can be computationally detected.
There are certain details that fall in between that can be managed with secondary services.
For example, information about the author's reporting history or retractions could be sourced from third-party databases, such as Muck Rack~\cite{MuckRack34online}.
Additional complexity may arise whenever subjective ratings for transparency cues have to be computationally scored.
For example, designers may seek to compute how well a news report fits with newsworthiness criteria, and this information could be subjective to the audience.
Crowdsourcing of such subjective ratings may be a possible solution to this issue.
Other works have shown promise in crowdsourcing subjective ratings (e.g., credibility)~\cite{bhuiyan2020investigating}.
}

\subsubsection{Considering news consumers' changing values}
\additionnew{
Apart from the specific design issues of production-side aspects, our study indicates that designers would have to take into account the changing habits of news consumers, such as increasing polarization and cynicism.
For such changing values, designers can personalize transparency cue design.
For instance, some news consumers (U3) preferred earlier journalistic norms (``Cronkite era'' reporting) such as objectivity; that is, reporting the facts only while leaving the interpretation and implications up to the audience.
For these news consumers seeking fact-only news, designers could focus on certain transparency features (e.g., presence of evidence cues) that highlight facts in an article.
On the other hand, there were news consumers (U1 and U9) who considered such an understanding of objectivity outdated~\cite{rosen1993beyond} and instead asked journalists to be open about their biases.
For those who asked for transparency on biases, designers could show features (e.g., conflict of interest cues) that disclose this information.
In either case, future work focusing on how each design affects trust for readers with varying political stances could provide further guides for designers, especially since prior scholarly works suggests that design interventions may or may not affect perceptions of trustworthiness depending on users' political ideologies~\cite{wood2019elusive,roozenbeek2019fake}.
%
}

\subsubsection{Resiliency considerations to avoid manipulation}
Some news consumers were concerned about the trustworthiness of transparency cues.
Considering that transparency features can also be exploited by bad actors, these features must be made resilient to manipulation.
In our study, each design cue has different levels of resilience to manipulation. 
While bad actors can easily manipulate numbers (e.g., an author's experience in years), it is harder to exploit cues that require further documentation with proof (e.g., for-/nonprofit organizations' legal documents, police reports as evidence). 
Several steps can be taken to further prevent exploitation of these cues.
Interdependent cues can be devised, such that manipulation of certain attributes can reveal inconsistencies or defects in the cues.
For example, the number of reports and percentage of corrections by an author have a dependency that can signal manipulation of each individually, but not in unison.
Resiliency can also be achieved through a third-party authority who verifies this data (e.g., an ombudsman and third-party fact-checkers). 
Eventually, frequent audits may also help keep the system in check, albeit at the cost of resources and time. 
However, it is important to note that bad actors with enough resources (e.g., state-funded actors) may still manipulate highly resilient features.

\subsubsection{Can transparency do harm?}
\additionnew{The suggestion that transparency can improve trust is quite complicated in practice.
As \additionminor{O'Neill} argues, despite increasing transparency, trust has receded~\cite{o2002question}.
According to \additionminor{her}, while transparency may help when there is prior deception, increasing transparency may also ``produce a flood of unsorted information and misinformation that provides little but confusion unless it can be sorted and assessed.''
As our journalists noted, some features could create a false equivalence between reporters from different organizations.
Given the level of skepticism in news media~\cite{1America53online}, more transparency could cause consumers to become more skeptical about news---perhaps even about factual reports.
To counter, our results offer particular priorities to consider (e.g., news selection and framing) in a structured fashion (e.g., through proper comparison across organizations).
There are also considerations for transparency around specific areas.
As Cunningham notes, ``To assume that we can know what someone thinks by identifying their personal traits, habits and predilections is a dangerous notion, and really has nothing to do with clarity.''~\cite{cunningham2006skin}
Our journalists remarked that transparency around sources could be harmful.
Some sources prefer to stay anonymous for a variety of reasons, such as preventing backlash from the organization.
Some journalists also prefer to keep sources anonymous out of necessity, so as not to lose potential sources of information on future events.
Given these constraints, future research could be directed towards quantifying the benefits and drawbacks of each transparency cue.
}

\subsection{\additionminor{Limitations}}
\additionnew{
In this study, we explore a subset of the larger issue that is growing distrust in the media; that is, how transparency through design cues can contribute to improving perceptions of trust for news articles.
Our work is limited in focus, as we prioritize the perceptions of two particular groups of stakeholders: news consumers and journalists.
While news organizations may play a much larger role in what gets adopted on their sites, our intent is to inform HCI designers and these organizations of the recommendations from our two stakeholder groups.
This work is also limited in its perspective on designing features, to the exclusion of examining and setting organizations' policies.}
\additionminor{It is worth noting that we did not consider journalists from alternative media (e.g., Breitbart, One America News Network, and Newsmax), who are known for their particular (and perceivably strong) partisan views. Journalists in such organizations with significant viewership may view these questions of trust and transparency differently, and they may need separate attention in future works.}

\additionnew{
Furthermore, while we used a scenario-based design to elicit feedback on which transparency cues can act as trust indicators, we can not definitively predict how the cues may affect users' trust without conducting a controlled user study on a functional system, given the ambiguous effect found in past experiments~\cite{karlsson2014you,curry2019effects}.
However, similarly to one of the experiments, the addition of multiple transparency features to a news item has the potential to affect users' perceptions of credibility~\cite{curry2019effects}.
Still, taking these design considerations into account, we have to be careful to test whether such a design could simply overwhelm news consumers, resulting in an undesirable opposite effect.
}

Result-wise, our study suffers from some limitations common in qualitative studies. First, we had a limited number of participants.
Furthermore, we could only interview people who agreed to participate.
Therefore, a self-selection bias exists in our study.
\additionnew{Moreover, we reached a majority of our participants by a combination of convenience and snowball sampling strategies.
Though these are acceptable strategies in social science research~\cite{bernstein2011trouble}, their use adds to the limitations of our study.
}
Even so, we tried to address this problem by reaching out to a variety of sources.
For example, in our news consumer pool, we had a mix of news-savvy and non-savvy participants with diverse backgrounds.
Additionally, as we progressed through our interviews, very few new themes emerged in the final interviews of each group.
Therefore, our results may have reached empirical saturation despite the limited sample size.
Second, our participant population is mainly US-based, with some exceptions; therefore, our findings might not be generalizable to other countries.
\additionnew{Indeed, trust in the media varies by country~\cite{HowMuchD88online}. Since we conducted the study during the latter years of the Trump administration, there might be some period-specific effects that influenced the perceptions of both groups of participants (e.g., the concern \additionminor{over} objectivity in news selection).}
Future studies in other regions and time periods might be able to address this limitation.

\section{Conclusion}
In this work, we examined how designers can adopt transparency features as indicators of trust on news websites. 
We explored these questions in a dual-perspective setting, interviewing journalists and news consumers with a scenario-based approach. 
Our results imply that HCI designers \additionnew{can offer indicators of trustworthiness} through transparency cues that reflect objectivity and evidence in news articles, authors' expertise, the process of reporting, and personal and organizational conflict of interest.
Both groups agreed on some cues while offering differing views on others.

\section{Acknowledgements}
\additionminor{This paper would not be possible without our participants. We would also like to thank the National Science Foundation for their support through grant \#$2128642$. We also appreciate valuable feedback from the members of the Social Computing Lab at Virginia Tech and University of Washington.}

\bibliographystyle{ACM-Reference-Format}
\bibliography{sample-base}
\received{October 2020}
\received[revised]{April 2021}
\received[accepted]{July 2021}

\newpage
\appendix
\section{Appendix}
\subsection{Interview Questions for the News Consumers}
\noindent\textbf{Demography}
\begin{itemize}
    \item Tell me about your background?
    \item How would you describe your political orientation?
    \item How often do you read news?
    \item What type of news do you read?
\end{itemize}

\noindent\textbf{Questions}
\begin{itemize}
    \item In your lifetime, do you think the news industry has changed? In what way?
    \item As you are probably aware, the subject of fake news is now a popular topic. Has this focus on fake news impacted your news consumption in any way?
    \item In the last couple of decades, Pew Research has found that public trust in mainstream news organizations has been declining. What do you think are the reason for that?
\end{itemize}

\begin{itemize}
    \item Some people have said that journalists can build trust by practicing more transparency in journalism? Transparency has been defined as building features into a news website that disclose how news sites collect, report and disseminate news. Do you think that newsrooms should be more transparent in these areas?
    \item If you do agree, what ways do you think journalists could improve their transparency.
    \item (Showing the design) Feature A contains an information box that provided details about the person who wrote the story such as the number of years they have been in journalism and whether they have had retractions. 
    As you know most news stories contain a ``who, what, when, where and why.'' Often times, fake news omits some of these characteristics. 
    In Feature B we could construct an algorithm (semi-supervised system) that highlights the presence or absence of these pieces of content. 
    Some stories, particularly fake news stories, contain indicators of bias or commentary. In Feature C, we provide indicators to show the degree to which the story follows Summary News Lead (SNL) or inverted pyramid-style reporting, fact-based reporting or opinion-based commentary. Would it be helpful to make these more visible to the public?
\end{itemize}


\begin{itemize}
    \item Would there be any other interactive type features that you would think could be beneficial?
\end{itemize}

\subsection{Interview Questions for the Journalists}
\noindent\textbf{Demography}
\begin{itemize}
    \item Tell me about your background?
    \item How often do you read news?
    \item What type of news do you read?
    \item Tell me about your position? What types of things do you do in your job?
    \item How long have you been in journalism?
\end{itemize}

\noindent\textbf{Questions}
\begin{itemize}
    \item What are some of the major changes that have happened in the field in your time?
    \item As you are probably aware, the subject of fake news is now a popular topic. Has this focus on fake news impacted you?
\end{itemize}

\begin{itemize}
    \item Some people have said that journalists can build trust by practicing more transparency in journalism? Transparency has been defined as building features into a news website that disclose how news sites collect, report and disseminate news. Do you think that newsrooms should be more transparent in these areas?
    \item If you do agree, what ways do you think journalists could improve their transparency.
\end{itemize}

\begin{itemize}
    \item (Showing the design) Feature A contains an information box that provided details about the person who wrote the story such as the number of years they have been in journalism and whether they have had retractions. 
    As you know most news stories contain a ``who, what, when, where and why.'' Often times, fake news omits some of these characteristics. 
    In Feature B we could construct an algorithm (semi-supervised system) that highlights the presence or absence of these pieces of content. 
    Some stories, particularly fake news stories, contain indicators of bias or commentary. In Feature C, we provide indicators to show the degree to which the story follows Summary News Lead (SNL) or inverted pyramid-style reporting, fact-based reporting or opinion-based commentary. Would it be helpful to make these more visible to the public?
\end{itemize}

\begin{itemize}
    \item Would there be any other interactive type features that you would think could be beneficial?
\end{itemize}

\end{document}